Physical Sciences-Applied Physical Sciences

# Gate-tunable carbon nanotube-MoS$_2$ heterojunction p-n diode

Deep Jariwala,[1] Vinod K. Sangwan,[1] Chung-Chiang Wu,[1] Pradyumna L. Prabhumirashi,[1] Michael L. Geier,[1] Tobin J. Marks,[1,2,*] Lincoln J. Lauhon,[1] and Mark C. Hersam[1,2,3,*]

[1]Department of Materials Science and Engineering, Northwestern University, Evanston, IL 60208, USA.

[2]Department of Chemistry, Northwestern University, Evanston, IL 60208, USA.

[3]Department of Medicine, Northwestern University, Evanston, IL 60208, USA.

*Corresponding authors. E-mail: t-marks@northwestern.edu and m-hersam@northwestern.edu

**Abstract:**

**The p-n junction diode and field-effect transistor (FET) are the two most ubiquitous building blocks of modern electronics and optoelectronics. In recent years, the emergence of reduced dimensionality materials has suggested that these components can be scaled down to atomic thicknesses. Although high performance field-effect devices have been achieved from monolayered materials and their heterostructures, a p-n heterojunction diode derived from ultrathin materials is notably absent and constrains the fabrication of complex electronic and optoelectronic circuits. Here, we demonstrate a gate-tunable p-n heterojunction diode using semiconducting single-walled carbon nanotubes (s-SWCNTs) and single-layer molybdenum disulfide (SL-MoS$_2$) as p-type and n-type semiconductors, respectively. The vertical stacking of these two direct band gap semiconductors forms a heterojunction with electrical characteristics that can be tuned with an applied gate bias to achieve a wide range of charge transport behavior ranging from insulating to rectifying with forward-to-reverse bias current ratios exceeding $10^4$. This heterojunction diode also responds strongly to optical irradiation with an external quantum efficiency (EQE) of 25% and fast photoresponse < 15 μs. Since SWCNTs have a diverse range of electrical properties as a function of chirality, and since an increasing number of atomically thin 2D nanomaterials are being isolated , the gate-tunable p-n heterojunction concept presented here should be widely generalizable to realize diverse ultrathin, high-performance electronics and optoelectronics.**

**KEYWORDS:** 2D transition metal dichalcogenides, single layer MoS$_2$, single walled carbon nanotubes, van der Waals heterostructure, p-n heterojunction, photodetectors

**Significance Statement:**

The p-n junction diode is the most ubiquitous and fundamental building block of modern electronics with far-reaching applications including integrated circuits, detectors, photovoltaics, and lasers. With the recent discovery and study of atomically thin materials, opportunities exist for adding new functionality to the p-n junction diode. Here, we demonstrate that a p-n heterojunction diode based on atomically thin $MoS_2$ and sorted semiconducting carbon nanotubes yields unprecedented gate tunability in both its electrical and optical properties, not observed in the case of bulk semiconductor devices. In addition to enabling advanced electronic and optoelectronic technologies, this p-n heterojunction diode provides new insight into charge transport and separation at atomically thin heterointerfaces.

\body

**Text:**

When two semiconductors with opposite carrier type contact one another, charge transfer occurs across their interface and creates a potential difference determined by the doping profile. In bulk semiconductor p-n junctions, the doping level is primarily controlled via diffusion or implantation of substitutional impurities, which implies minimal control over the doping profile following device fabrication. In contrast, atomically thin semiconductors can be electrostatically doped by applying a bias to a capacitively coupled gate electrode.(1-3) The atomically thin structure of these materials also enables doping modulation of the overlying layers in a vertically stacked heterostructure(4). For example, this strategy allows gapless graphene to be used in field-effect tunneling devices in combination with other layered materials(4, 5). Vertical 2D heterostructures have also been used to create high-performance MOSFETs(6), tunneling FETs(4), barristors(7), inverters(8), and memory devices(9, 10), in addition to facilitating the study of novel physical phenomena in layered materials(11-14). Similarly, in-plane graphene heterostructures and controlled doping have served as the basis for unique 2D devices(15-18). Although the nearly perfect 2D structure and low density of states in graphene provide advantages in some heterostructure devices, its gapless nature prevents the formation of a large potential barrier for charge separation and current rectification despite efforts to create in-plane p-n homojunctions by split gating(19). In particular, the lack of distinct monolayer semiconductors with complementary (p and n) polarities has precluded the realization of a gate-tunable heterojunction p-n diode.

In this report, we demonstrate the fabrication and operation of a gate-tunable p-n heterojunction diode using s-SWCNTs (p-type)(20) and SL-MoS$_2$ (n-type)(21, 22). Figure 1A shows a false-colored scanning electron micrograph of a representative device. A lower magnification optical micrograph of the same device in Figure 1B shows the n-type (SL-MoS$_2$) FET, p-n s-SWCNT/SL-MoS$_2$ heterojunction, and p-type (s-SWCNT) FET from top to bottom. The device fabrication begins by depositing and e-beam lithographically defining gold electrodes on SL-MoS$_2$ flakes that are exfoliated on 300 nm SiO$_2$/Si wafers (Figure 1C, I). The SL-MoS$_2$ FET is electrically isolated by patterning 30 nm alumina *via* atomic layer deposition (ALD) followed by transfer and patterning of a sorted s-SWCNT thin film(23, 24) (Figure 1C, II) to yield the final device (Figure 1C, III). Supplementary Information Section 1 provides further details on the device fabrication.

The ultrathin nature of the heterojunction allows gate tunability of the diode electrical characteristics. Figure 2A shows the output plots of a representative device under different gate biases. The device transitions from a nearly insulating behavior at $V_G$ = 70 V to a poorly rectifying state ($r^2$ < 60% for fits to the Shockley diode equation) at $V_G$ = 40 V to a highly rectifying diode for $V_G \leq 0$ V ($r^2$ > 99%). For the heterojunction diodes, $V_D$ refers to the bias on the s-SWCNT electrode such that $V_D > 0$ corresponds to forward bias while the electrode in contact with SL-MoS$_2$ is grounded. The transfer plot further demonstrates the gate tunability of the current through the p-n heterojunction (Figure 2B). The transfer characteristics of the junction (green) show an unusual gate voltage dependence, which we refer to as '*anti-ambipolar*' behavior. In particular, the maximum conductance occurs near $V_G$ = 0, which is the opposite of conventional ambipolar behavior that shows a minimum conductance near $V_G$ = 0. The current on/off ratio exceeds $10^4$ in the transfer plots, which is suitable for advanced logic applications (see Supplementary Information Section 3a). The transfer characteristics of the heterojunction can be qualitatively viewed as a superposition of the p-type s-SWCNT (red) and n-type SL-MoS$_2$ (blue) FET transfer plots. However, the net current through the heterojunction is smaller than the SL-MoS$_2$ and s-SWCNT FET transfer characteristics due to the additional resistance at the junction.

This unique transfer characteristic can be viewed as originating from an FET channel consisting of two p and n semiconductors in series. The change in the resistance of each component with gate bias affects the net series resistance leading to the resulting anti-ambipolar transfer characteristic of the junction. Thus the junction transfer curve has two off states, representing the depleted state of SL-MoS$_2$ and s-SWCNTs. Furthermore, the contact resistances of the s-SWCNT film and SL-MoS$_2$ with Au are relatively small compared to the sheet/channel resistances (25, 26) and thus have negligible effect on the junction characteristics. (See Supplementary Information Section 3b for output plots of s-SWCNT and SL-MoS$_2$ FETs). It is worth noting that the maximum of the junction transfer curve does not exactly coincide with the intersection of the SL-MoS$_2$ and s-SWCNT transfer curves. This offset can be attributed to hysteresis in the transfer plots and the fact that the SL-MoS$_2$ FET is encapsulated in alumina, which is known to shift threshold voltage in SL-MoS$_2$ FETs.(21)

An important parameter in characterizing diode characteristics is the ratio of the forward current, $|I_f|$, to the reverse current, $|I_r|$, at the same bias magnitude. Figure 2C shows that $|I_f|/|I_r|$ varies by over five orders of magnitude as a function of gate voltage. Between the two gate bias extremes, the s-SWCNT/SL-MoS$_2$ heterojunction transitions from an n-n$^+$ junction at $V_G$ = 80 V to a p-n junction at $V_G$ = -80 V. The large band gap of SL-MoS$_2$ (> 1.8 eV)(27) compared to that of the s-SWCNTs (~0.7 eV) allows electrostatic depletion of SL-MoS$_2$ to a lightly n doped (n$^-$) or nearly intrinsic insulating state, thereby leading to $|I_f|/|I_r|$ values exceeding 10$^4$ at $V_G$ = -80 V. On the other hand, the small band gap of the s-SWCNTs allows electrostatic inversion from p-doping to n-doping at large positive $V_G$, resulting in poor $|I_f|/|I_r|$ values for $V_G$ > 60 V. Gate tunable p-n homojunction diodes have been previously fabricated by split gating of individual SWCNTs. However, such homojunctions only allow control over the built-in voltage via differential biases in split gates. (28, 29). The present gate tunable p-n heterojunction, on the other hand, is fundamentally different as it has a built-in potential at zero gate bias as evident from the rectifying I-V characteristics (Figure 2A). Furthermore, in this heterojunction, the gate is used to simultaneously tune the doping concentrations of both semiconductors, thereby allowing tunability in the built-in voltage and rectification ratios.

To further understand the gate-dependent modulation of the heterojunction I-V curves, we fit them to the Shockley diode equation. The best fit to the Shockley diode equation is observed for $V_G$ = -40 V. For other $V_G$ values, either the diode ideality factor ($n$) is > 2 or the fits are poor ($r^2$ < 60%) (see Supplementary Information Section 3c). The disorder at the interface of random network SWCNT films (see AFM images in Supplementary Information Section 1c) and SL-MoS$_2$ possibly leads to more recombination, resulting in larger ideality factors as compared to the nearly ideal diode behavior in WSe$_2$/InAs heterojunctions that have a uniform 2D interface(30). The gate dependence of the present diode behavior enables gate-tunable rectifier circuits (Figure 2D), which is a unique feature of the s-SWCNT/SL-MoS$_2$ heterojunction device that has not been observed in conventional bulk semiconductor diodes (see Supplementary Information Section 2 for further details on measurement techniques).

Both SL-MoS$_2$ and s-SWCNTs have direct band gaps(31, 32) and exhibit signatures of bound excitonic states in their absorption spectra.(2, 31) Therefore, photocurrent generation is expected upon optical irradiation of p-n heterojunctions based on these materials. To that end, scanning photocurrent microscopy was employed to spatially map the local photoresponse of the s-SWCNT/SL-MoS$_2$ heterojunction device (Figure 3A). The regions of large negative photocurrent lie in the heterojunction area outlined by the SL-MoS$_2$ flake (purple) overlapping with the patterned s-SWCNT film (red). No measurable photocurrent is observed from the non-overlapping regions of either the s-SWCNT film, SL-MoS$_2$, or the electrical contacts (see Supplementary Information Section 4 for additional details), indicating that the photocurrent measured under uniform illumination is generated by the vertical heterojunction. Photocurrent from Au contacts to s-SWCNT films was not observed in our recent study(33). For the case of two-terminal SL-MoS$_2$ devices, photocurrent has been observed at the contacts due to band bending(34). However, the band-bending at the contacts is opposite to that of the junction, which suppresses the local near-contact photocurrent because charge neutrality cannot be maintained at the injection level used. The spectral dependence of the photocurrent (Figure 3B) corresponds to the absorption peaks of SL-MoS$_2$(27) and S$_{22}$(24) peaks of s-SWCNTs, which demonstrates that this novel heterointerface can induce carrier separation following exciton and/or free carrier generation in either material. The photocurrent generated in the visible portion of the spectrum also likely has contributions from both SL-MoS$_2$ and the S$_{33}$ transitions of s-SWCNTs since both absorb in that range of wavelengths (see Supplementary Information Figure S2).

To further illustrate the photoresponse of the s-SWCNT/SL-MoS$_2$ heterojunction, both output (I-V) and transfer curves (I-V$_g$) were acquired under global illumination at a series of wavelengths. A representative comparison of the dark and illuminated I-V curves at V$_G$ = -40 V reveals that the photocurrent increases by 4 orders of magnitude at a heterojunction reverse bias of -5 V (Figure 3C). Figure 3D also shows the gate voltage dependent photocurrent values at a heterojunction reverse bias of -10 V (Figure 3D). As the gate voltage becomes more negative, the relative contribution from the s-SWCNT portion of the spectrum decreases, which is consistent with the s-SWCNT/SL-MoS$_2$ heterojunction becoming a p$^+$-n$^-$ junction. Since the the depletion region in a p-n junction extends farther into the side with lower doping/majority carrier concentration, the junction almost entirely lies in the SL-MoS$_2$ as it is depleted at negative V$_G$, leading to a reduced photocurrent contribution from the s-SWCNTs as observed in Figure 3D. If we consider the junction from a molecular perspective used to describe organic semiconductor heterojunctions, one might also expect a change in the rate of charge transfer due to changes in band offsets (or HOMO and LUMO levels) across the heterojunction.

With a strong photoresponse, the s-SWCNT/SL-MoS$_2$ heterojunction can be exploited as a photodetector. Diode-based photodetectors are known for their fast photoresponse times compared to phototransistors since the photoexcited carriers must only traverse a distance equal to the depletion width of the junction. Indeed, we observe fast photoresponse (< 15 µs) from the s-SWCNT/SL-MoS$_2$ heterojunction as seen in Figure 4A,B. Importantly, this photoresponse time is orders of magnitude smaller than recently reported heterojunction phototransistors based on graphene(35). Furthermore, the s-SWCNT/SL-MoS$_2$ heterojunction photoresponse time reported here is limited by the rise time of the preamplifier used in these experiments, and thus 15 µs represents an upper bound.

The external quantum efficiency (EQE) of the s-SWCNT/SL-MoS$_2$ heterojunction photodetector is also noteworthy (~25% at V$_D$ = -10 V) and compares favorably to other recently reported 2D nanomaterial heterostructures (Figure 4C) (35-37). The EQE is calculated as $EQE = \frac{I_{ph}/e}{P/h\nu} \times 100$, where I$_{ph}$, e, P, and hν represent the photocurrent, electronic charge, incident optical power, and photon energy, respectively. The linear rise in EQE at low reverse biases (Figure 4C) is similar to the behavior of quantum dot (QD) based photodiodes,(38) which is related to changes in the depletion region. Here, given the negligible thickness of SL-MoS$_2$ compared to the expected depletion width(39), we attribute the increasing EQE to an increase in the rate of charge transfer across the heterojunction interface due to the change in band alignment, rather than to an increase in the depletion width. The spectral responsivity (*R*) of the heterojunction photodetector is presented in Figure 4D. The highest *R* exceeds 0.1 A/W at a wavelength of 650 nm, which is comparable to other nanostructured diode-based photodetectors in the literature(40). The combination of high *R* and fast photoresponse time presents distinct advantages over currently available organic and QD photodiodes(38).

In conclusion, we have demonstrated a gate-tunable p-n heterojunction diode through the integration of p-type s-SWCNTs and n-type SL-MoS$_2$. The ultrathin nature of the constituent materials implies that both components can be modulated by a capacitively coupled gate bias, thereby enabling wide tunability of charge transport from a nearly insulating state to a highly rectifying condition with forward-to-reverse bias current ratios exceeding 10$^4$. When operated as a three-terminal device, the p-n heterojunction diode also shows 'anti-ambipolar' behavior with current on/off ratios greater than 10$^4$, suggesting its utility in advanced logic applications.

Furthermore, since s-SWCNTs and SL-MoS$_2$ are direct band gap semiconductors, the p-n heterojunction diode serves as an effective photodetector with fast photoresponse < 15 µs. By combining other chirality-resolved s-SWCNTs(41) with the growing list of 2D semiconductor nanomaterials(42-45), the p-n heterojunction diode can be generalized to a wide range of electronic and optoelectronic applications.

**Acknowledgements:** This research was supported by the Materials Research Science and Engineering Center (MRSEC) of Northwestern University (NSF DMR-1121262) and the Office of Naval Research MURI Program (N00014-11-1-0690). The authors acknowledge discussions with J. M. P. Alaboson, and also thank B. Myers and I. S. Kim for assistance with electron beam lithography and Raman spectroscopy, respectively. M.L.G. also acknowledges a National Science Foundation Graduate Research Fellowship. This research made use of the NUANCE Center at Northwestern University, which is supported by NSF-NSEC, NSF-MRSEC, Keck Foundation, and State of Illinois. This research also utilized the NUFAB cleanroom facility at Northwestern University.

**Author Contributions:** D.J, V.K.S, and M.C.H conceived the device structure and experiments. D.J designed and fabricated the device with assistance from V.K.S. D.J, V.K.S, and C.C.W designed and performed the electrical and photocurrent measurements. P.L.P and M.L.G sorted the SWCNTs and carried out the SEM and AFM characterizations. T.J.M., L.J.L., and M.C.H. oversaw the research, and all authors participated in the data analysis and writing of the manuscript.

**Additional Information:** The authors declare no competing financial interests. This article contains supplementary information online at www.pnas.org/lookup/suppl/doi:xxx. Correspondence and requests for materials should be addressed to M.C.H (m-hersam@northwestern.edu).

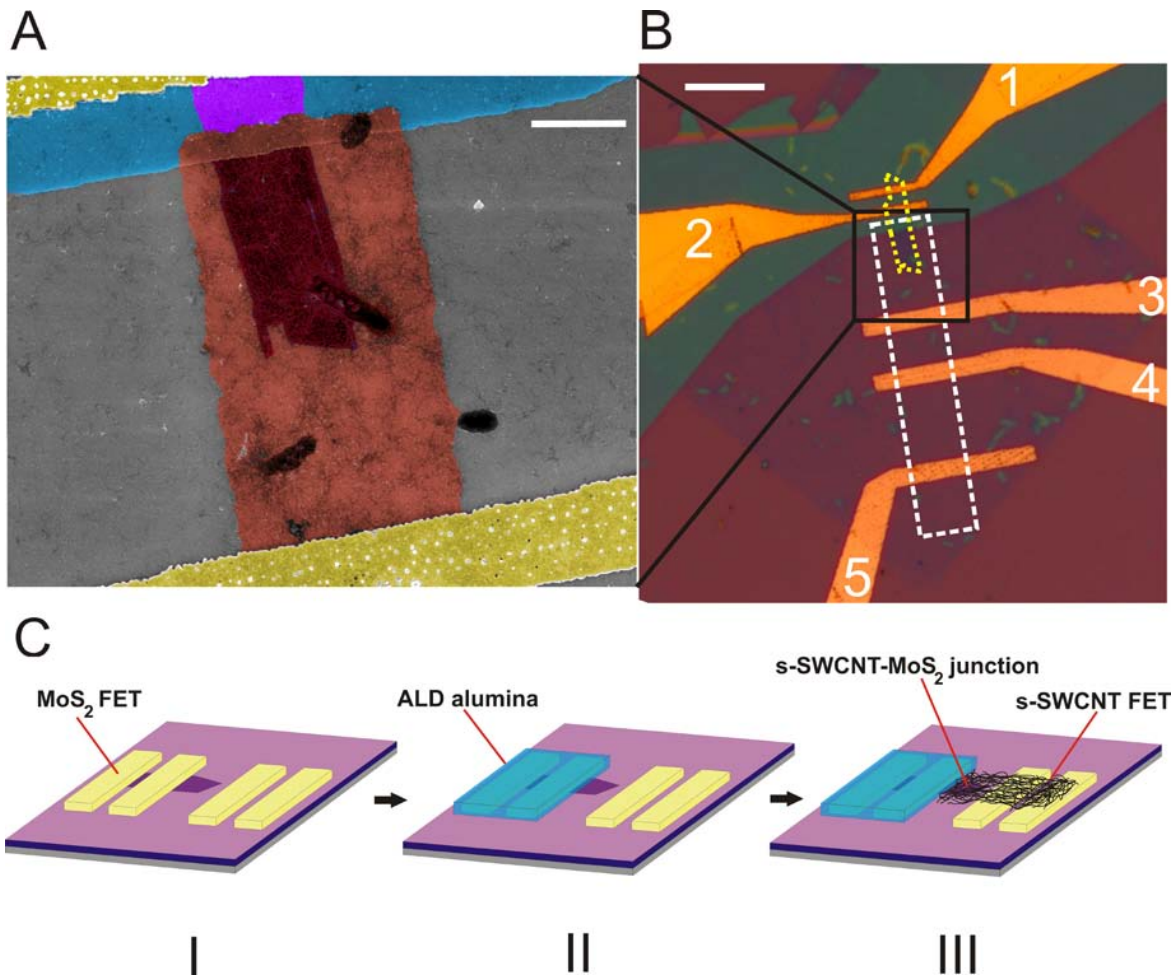

**Figure 1: Microscopy and fabrication of the s-SWCNTs/SL-MoS₂ p-n heterojunction diode. A,** False-colored scanning electron microscopy (SEM) image of the heterojunction diode (scale bar = 2.5 μm). The yellow regions at the top and bottom are the gold electrodes. The patterned alumina (blue region) serves as a mask for insulating a portion of the SL-MoS₂ flake (violet region). The pink region is the patterned random network of s-SWCNTs (p-type) in direct contact with the exposed part of SL-MoS₂ flake (n-type) to form the p-n heterojunction diode (dark red). **B,** Optical micrograph showing the device layout at a lower magnification. The dashed yellow boundary indicates the SL-MoS$_2$ flake while the dashed white rectangle denotes the patterned s-SWCNT film. Electrodes 1 and 2 form the n-type (SL-MoS$_2$) field-effect transistor (FET), which is insulated by the patterned alumina film (cyan). Electrodes 2-3 form the p-n heterojunction while 3-4 and 4-5 form p-type s-SWCNT FETs. (scale bar = 10 µm). **C,** Schematic of the fabrication process: (I) SL-MoS$_2$ FET and an extra pair of electrodes are fabricated *via* e- beam lithography on 300 nm SiO$_2$/Si. The Si substrate acts as the global back gate. (II) The MoS$_2$ FET is insulated by patterning an alumina film in a liftoff process, followed by (III) transfer and patterning of the s-SWCNT network to yield the final device configuration consisting of a top contact SL-MoS$_2$ FET, bottom contact s-SWCNT FET, and p-n heterojunction.

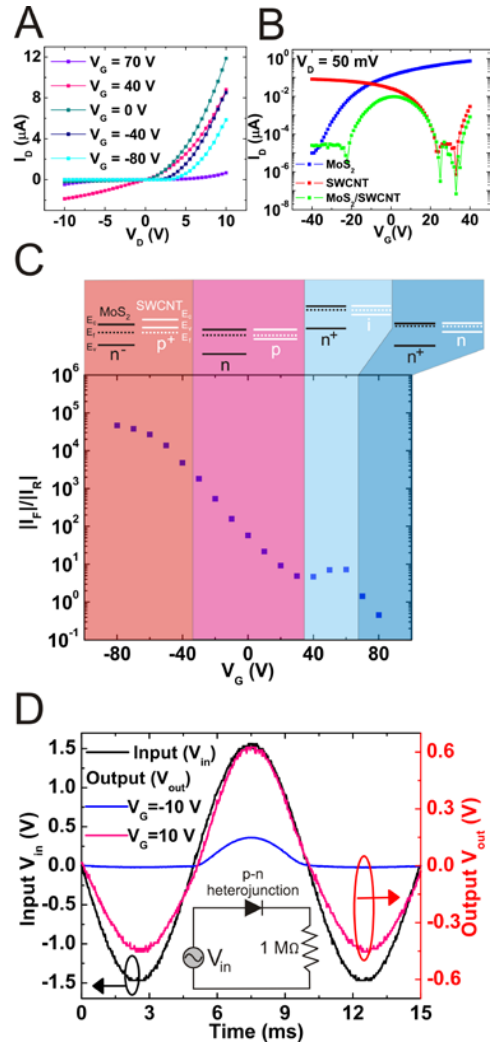

**Figure 2: Electrical properties of the s-SWCNT/SL-MoS$_2$ p-n heterojunction diode. A,** Gate-tunable output characteristics showing the transition from a nearly insulating state at $V_G$ = 70 V to a conductive state with relatively poor rectification at $V_G$ = 40 V to a highly rectifying diode behavior at negative gate voltages. **B,** Transfer characteristics of the p-n junction (green), showing an 'anti-ambipolar' characteristic, which is qualitatively a superposition of the transfer characteristics of the p-type s-SWCNT FET and n-type SL-MoS$_2$ FET. **C,** Forward-to-reverse current ratio (at a heterojunction bias magnitude of 10 V) as a function of gate bias. The labels

at the top show the corresponding band diagrams for the s-SWCNT/SL-MoS$_2$ p-n heterojunction. At a high positive gate bias, the formation of an n$^+$-n junction implies a low rectification ratio that transitions into an n$^+$-i junction (plateau region in the plot) with reducing $V_G$. The rectification ratio then rises with decreasing gate bias due to the formation of a p-n junction.**D,** Demonstration of gate-tunable rectification using the p-n heterojunction diode. The y-axis on the left shows the input voltage while the y-axis on the right shows the output voltage across the series resistor (1 MΩ). As a function of the gate bias, the device evolves from a non-rectifying resistor-like state at $V_G$ = 10 V (magenta) to a diode-like rectifying state at $V_G$ = -10 V (blue).

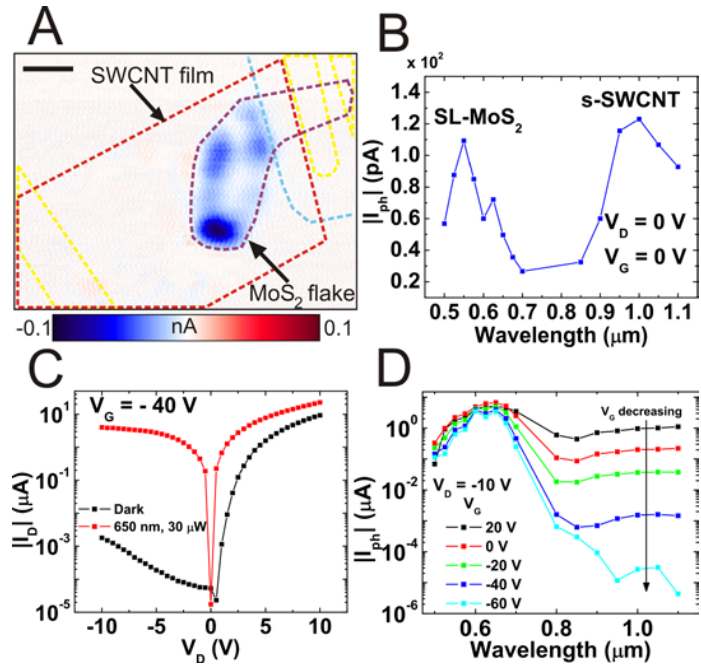

**Figure 3: Photoresponse of the p-n heterojunction. A,** Scanning photocurrent micrograph of a representative heterojunction device acquired at $V_D$ (s-SWCNT electrode), $V_G$ = 0 V showing the outlines of the SL-MoS$_2$ flake (purple dashed line) and the patterned s-SWCNT film (red dashed line) acquired at 700 nm with 20 µW power. Regions of large negative photocurrent (blue) are observed in the overlapping junction region. The patterned alumina and electrodes are indicated by cyan and yellow dashed lines, respectively. **B,** Photocurrent spectrum of the junction under global illumination and zero bias conditions. The photocurrent magnitude is highest at the characteristic absorption energies of both SL-MoS$_2$ and s-SWCNTs. The photocurrent spectrum is acquired at the same incident power (30 µW) **C,** Output curve of the same device in the dark and under global illumination at 650 nm. **D,** The photocurrent spectral response can be tuned with the gate voltage. With decreasing gate voltage, the increased p-

doping of the nanotubes and concomitant decreased n-doping of MoS$_2$ leads to a lower photocurrent in the near infrared region.

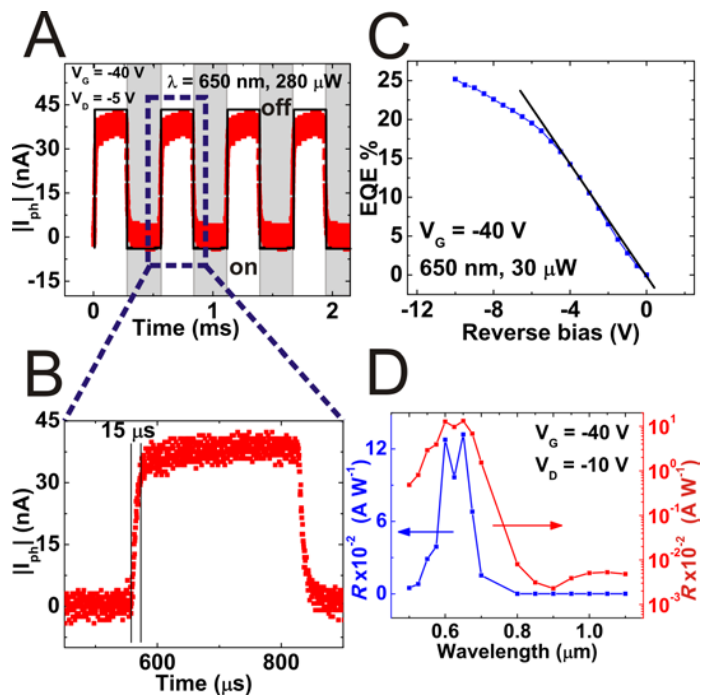

**Figure 4: Photodetection using the p-n heterojunction diode. A,B,** Time dependent photoresponse of the p-n heterojunction showing fast rise and decay times of ~15 μs. **C,** External quantum efficiency (EQE) as a function of reverse bias for the heterojunction at 650 nm. EQE increases linearly with reverse bias from 0 to -5 V with the highest EQE of 25% occurring at -10 V. **D,** Spectrally dependent responsivity (*R*) of the photodiode in linear (blue) and logarithmic (red) scales. A large responsivity is observed for the absorption wavelengths of SL-MoS$_2$ as compared to s-SWCNTs since the diode is being operated at $V_G$ = -40 V (depletion mode of SL-MoS$_2$).

Supplementary Information

# Gate-tunable carbon nanotube-MoS$_2$ heterojunction p-n diode

Deep Jariwala,[1] Vinod K. Sangwan,[1] Chung Chiang Wu,[1] Pradyumna L. Prabhumirashi,[1] Michael L. Geier,[1] Tobin J. Marks,[1,2,*] Lincoln J. Lauhon,[1] and Mark C. Hersam[1,2,3]*

[1]Department of Materials Science and Engineering, Northwestern University, Evanston, IL 60208, USA.
[2]Department of Chemistry, Northwestern University, Evanston, IL 60208, USA.
[3]Department of Medicine, Northwestern University, Evanston, IL 60208, USA.

* Corresponding author. E-mail: t-marks@northwestern.edu and m-hersam@northwestern.edu

## Section 1. Materials and Device Fabrication

### a. Raman Spectroscopy of MoS$_2$

The monolayer character of the MoS$_2$ flake samples was confirmed via Raman spectroscopy. The Raman spectra were acquired using a 532 nm laser with a 100x objective (NA = 0.9) in a scanning confocal microscope (WITec Alpha300 R). The separation between E$_{2g}$ and A$_{1g}$ modes of MoS$_2$ (Δ) is a well-known parameter for identifying layer thickness in ultrathin MoS$_2$ flakes. Typically a Δ value of < 20 cm$^{-1}$ indicates a single layer sample(1). Figure S1 shows the Raman spectrum (Δ~17.7 cm$^{-1}$) of a representative flake that was fabricated into a p-n heterojunction device.

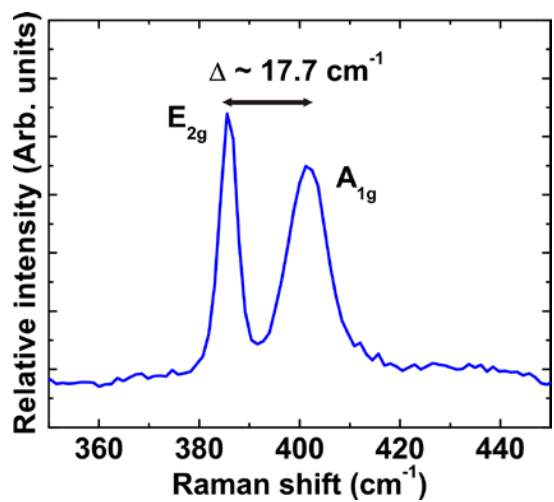

**Figure S1:** Raman spectrum of a representative $MoS_2$ flake. The separation of 17.7 cm$^{-1}$ between the $E_{2g}$ and $A_{1g}$ features indicates monolayer thickness.

b. Sorting and Deposition of Semiconducting Single-Walled Carbon Nanotubes

Semiconducting single-walled carbon nanotubes (SWCNTs) with ~99% semiconductor purity were isolated by density gradient ultracentrifugation following the procedure discussed in a previous report(2). SWCNT thin films were prepared by vacuum filtration and transferred onto pre-patterned Au electrodes by the acetone bath transfer method as outlined in the literature(2, 3). The semiconducting purity of the sorted SWCNTs was estimated using the optical absorbance characterization and an analysis protocol developed earlier(3, 4). Figure S2 shows the optical absorbance spectra for as-purchased arc-discharge SWCNTs (P2, Carbon Solutions Inc.) and sorted ~99% semiconductor purity SWCNTs. The average diameter of the semiconducting tubes is 1.4 nm.

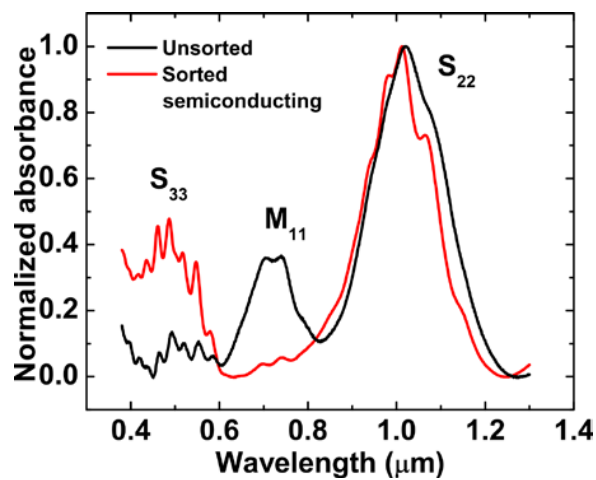

**Figure S2:** Normalized optical absorbance spectra of arc discharge SWCNTs. The unsorted SWCNTs (black) consist of a mixture of metallic and semiconducting species as evidenced by the presence of both semiconducting ($S_{22}$ and $S_{33}$) and metallic ($M_{11}$) peaks, while the spectrum of the sorted semiconducting SWCNTs (red) shows strong semiconducting peaks and negligible metallic peaks.

### c. Device Fabrication

All devices were fabricated on 300 nm thick $SiO_2$/Si substrates. The Si <100> wafers were purchased from Silicon Quest International. The wafers were doped n-type with As (resistivity = 0.001-0.005 Ω-cm). The $MoS_2$ FETs were fabricated using previously reported techniques.(5) Specifically, bulk $MoS_2$ crystals were purchased from SPI Supplies and mechanically exfoliated using Scotch tape. The flakes were identified using an optical microscope (Olympus BX 51M) and then subjected to electron-beam lithography (EBL). A two-step EBL process was adopted to fabricate Au contacts with no adhesion layer. The patterns in the first step were designed to be just short of touching the $MoS_2$ flakes. Au (75 nm thick) is the metal in contact with the $MoS_2$ flake. Following liftoff in acetone, the devices were further cleaned with remover PG (Microchem) at 60 °C for 1 hr.

A portion of the MoS₂ FET including both the contacts and channel was exposed in a subsequent step of EBL. Atomic layer deposition (ALD) (Cambridge Nanotech, Savannah S100) was used to grow 30 nm alumina at 100 °C to insulate the $MoS_2$ channel and contacts, thus prevent shorting following subsequent SWCNT deposition. Trimethyl aluminum (TMA) (Aldrich, 99%) was used as the precursor for ALD growth of alumina, and ultrahigh purity nitrogen (Airgas) was used as the purging gas. A single ALD cycle consisted of a TMA pulse for 0.015 s and a 30 s purge, followed by a $H_2O$ pulse for 0.015 s and a second 30 s purge. The growth rate was determined to be ~0.75 Å/cycle. A total of 400 cycles was performed to achieve 30 nm thickness. The oxide was lifted off in warm acetone at 50 °C for 1 hr. A cellulose membrane containing the semiconducting SWCNT film was then stamped onto the entire substrate and dissolved under acetone vapor as outlined in previous reports(2). The SWCNT film was annealed in vacuum (50 mTorr) at 200 °C for 1 hr and then patterned with EBL using a ZEP 150 (Nippon Zeon) resist. Reactive ion etching (Samco RIE-10 NR) in an oxygen plasma atmosphere (100 mW, 15 sec, 20 sccm) was then used to etch the nanotubes. The resist was subsequently dissolved in hot (80° C) N-methyl-2- pyrrolidone for ~6 hr. The SWCNT film morphology close to the junction region was characterized using atomic force microscopy (AFM) as shown in Figure S3.

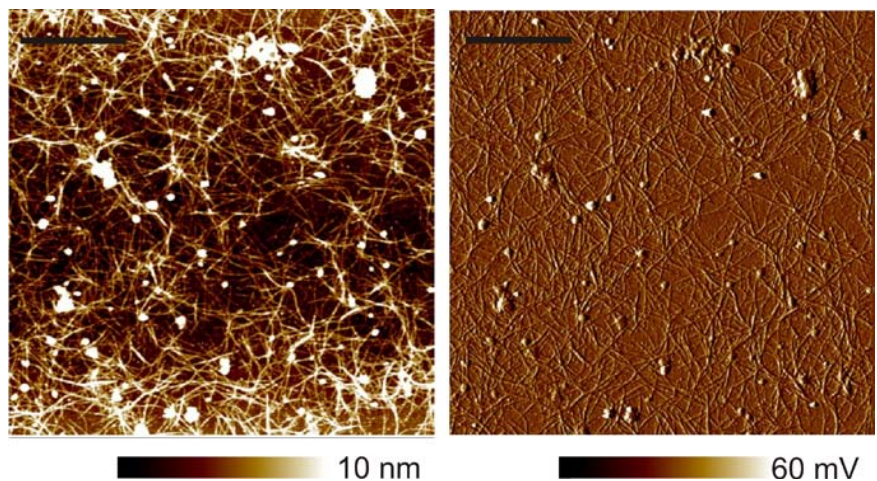

**Figure S3:** AFM images of the SWCNT film. Topographic image (left) and amplitude error (right) show the random network morphology and uniform coverage of SWCNTs (scale bar = 1.5 μm).

**Section 2. Electrical and Photocurrent Measurements**

Output and transfer characteristics were measured using Keithley 2400 source meters and custom LabView programs. The gate voltage was swept at 10 V/sec in steps of 1 V in the transfer and output plots shown in Figure 2B of the manuscript.

The p-n heterojunction diode was used as a half-wave rectifier using the circuit outlined in the inset of Figure 2D. A 1 MΩ resistor was used in series with the device to limit the current. A sinusoidal wave from a waveform generator was used as the input, while the output current was measured using a pre-amplifier (1211 DL Instruments) and the output voltage as it would appear across a 1 MΩ resistor (Schematic in Figure 2D). Time domain waveforms were captured using an oscilloscope.

A scanning confocal microscope (100x objective with NA = 0.9, WiTec system) coupled to a tunable coherent white light source (NKT Photonics) was used to generate the spatially resolved photocurrent, which was converted into a voltage by a current preamplifier and recorded by either a lock-in amplifier (for imaging) or a digital sampling oscilloscope (for temporally resolved measurements). The junction area was subjected to global illumination using the same apparatus with a 20x objective. The I-V characteristics under global illumination were acquired using the same Keithley 2400 source meters and custom LabView programs.

## Section 3

### a. Transfer Characteristics of the p-n Heterojunction Diode

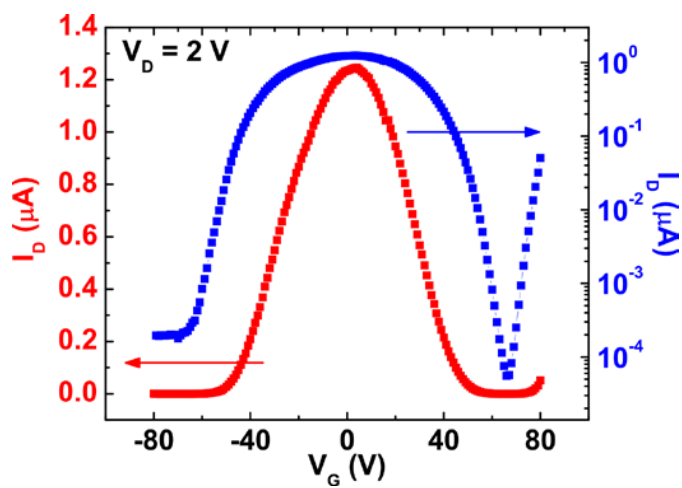

**Figure S4:** Linear (red) and semi-log (blue) transfer characteristics of a representative p-n heterojunction diode showing 'anti-ambipolar' behavior and on/off ratio exceeding $10^4$.

### b. Output Characteristics, Contact Resistance, and Sheet Resistance

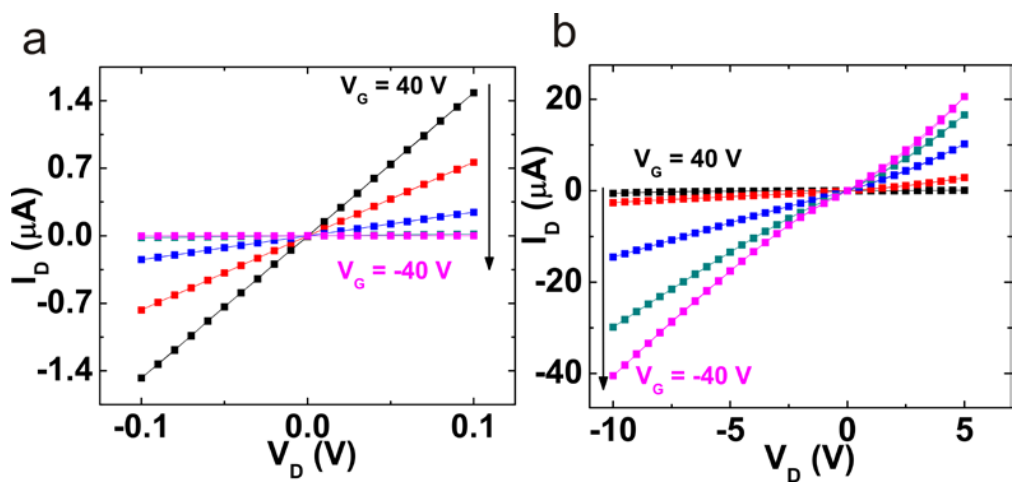

**Figure S5:** a, Output plots of a representative SL-MoS$_2$ FET showing linear I-V characteristics and n-type behavior. b, Output plots of a representative s-SWCNT FET showing p-type behavior.

The channel resistances were estimated at $V_G = 0$ V by subtracting the contact resistance values reported in literature from the $(dI/dV)^{-1}$ values obtained using the I-V plots shown above.

The channel resistance of the CNT FET at $V_G$ = 0 V is estimated to be approximately 0.66 MΩ while that of the MoS$_2$ FET is approximately 0.39 MΩ. These channel resistance values are > 15 times smaller than the junction resistance ($(dI/dV)^{-1}$ near $V_D$ = 0 V) of approximately 11.6 MΩ obtained from Figure 2A of the manuscript. Assuming that the length of the non-junction portions of SL-MoS$_2$ and s-SWCNT film are the same as their respective FET channels, the resistance of the junction after subtracting the series resistances of the non-junction portions and the contact resistances is about 10.4 MΩ. Thus, the junction resistance dominates even in presence of series resistance at $V_G$ = 0. However, this situation changes under depletion gate biases of either the s-SWCNTs or MoS$_2$, which leads to the anti-ambipolar behavior with two off states at either extremes in gate voltage.

### c. Data Fitting to the Shockley Equation

The diode output curves with varying gate bias ($V_G$) were fit with the Shockley diode equation, $|I_D| = |I_{rs}|(e^{\frac{eV_D}{nk_BT}} - 1)$, where $I_D$ is the drain current, $V_D$ is the drain bias, $I_{rs}$ is the reverse saturation current at $V_D$ = -0.05 V, e is the electronic charge, n is the ideality factor, $k_B$ is the Boltzmann constant, and T is the temperature of operation (300 K). The best fit to the diode curve ($r^2$>99%) is achieved for $V_G$ values ranging from -30 to -60 V with the diode ideality factor (n) approaching closest to 1 at -40 V (Figure S6). The I-V curves at other gate biases fit poorly to the Shockley equation ($r^2$<70 %).

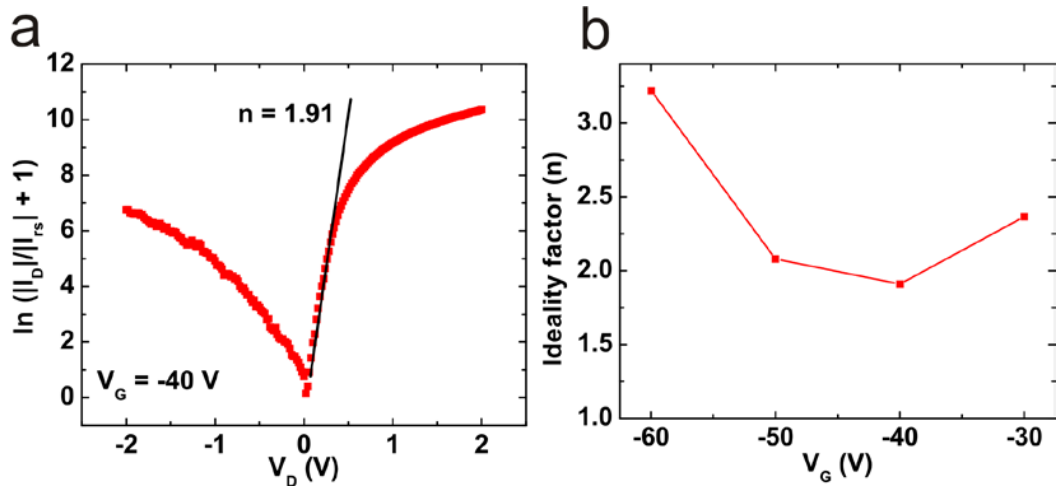

**Figure S6: a,** Shockley diode equation fit to the output curve at $V_G = -40$ V. **b,** The variation of n with $V_G$ shows that the diode is closest to ideal behavior at $V_G = -40$ V.

### Section 4. Spatial Mapping of Photoluminescence and Raman Shift

Spatial Raman and photoluminescence (PL) mapping was performed on the heterojunction device. A spatial map of the Raman 2G peak (3100-3250 cm$^{-1}$) from the SWCNTs and photoluminescence from the SL-MoS$_2$ at 3700-3900 cm$^{-1}$ (A peak) shows that the photocurrent arises only from the junction region. It was also observed that that the PL signal was uniform in the junction area of the MoS$_2$ flake as compared to the protected/masked area as seen in Figure S7.

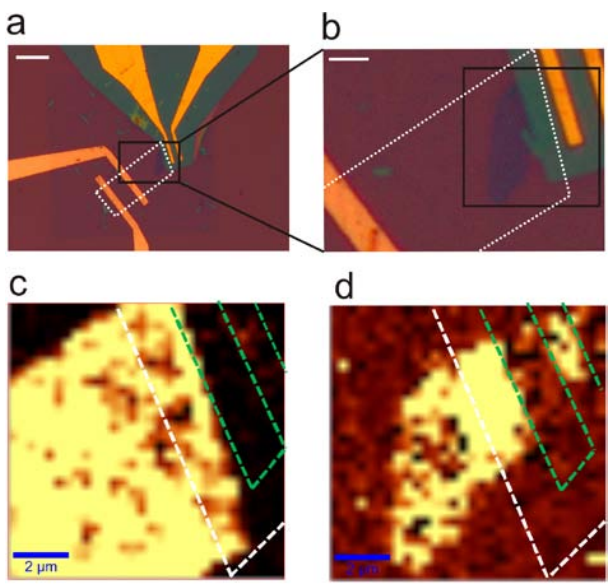

**Figure S7: a,** Optical micrograph of a heterojunction device showing a SL-MoS$_2$ FET top right and s-SWCNT FET bottom left. The white dotted line indicates the extent of the patterned SWCNT film (scale bar = 10 μm). **b,** Zoom-in of (a) representing the area scanned for spatial photocurrent mapping in Figure 3A of manuscript. The black outline indicates the scanned area for spatial mapping of the PL and the Raman shift (scale bar = 2.5 μm) **c,** Spatial map of the Raman shift (3100-3250 cm$^{-1}$), which shows the bright area as the patterned SWCNT film. No Raman signal for the SWCNTs is observed in the MoS$_2$ contact or channel region marked by green lines. The white boundary indicates the extent of ALD grown alumina. **d,** PL map of the SL-MoS$_2$ flake at the A peak (3700-3900 cm$^{-1}$). The region of increased PL intensity represents the shape of the flake as seen in the optical images and the photocurrent map of Figure 3A. No PL signal is observed in the contact regions, indicating PL quenching by the gold. MoS$_2$ PL is also observed from the SL-MoS$_2$ FET channel area, between the contacts.

**Section 5**

**a. Power Dependence of Photocurrent**

The power dependence of the photocurrent was measured under zero bias using the same illumination system. The power dependence is sub-linear which is consistent with bimolecular recombination of electrons and holes, further implying that the concentration of photogenerated carriers is similar to or greater than the intrinsic carrier concentration(6).

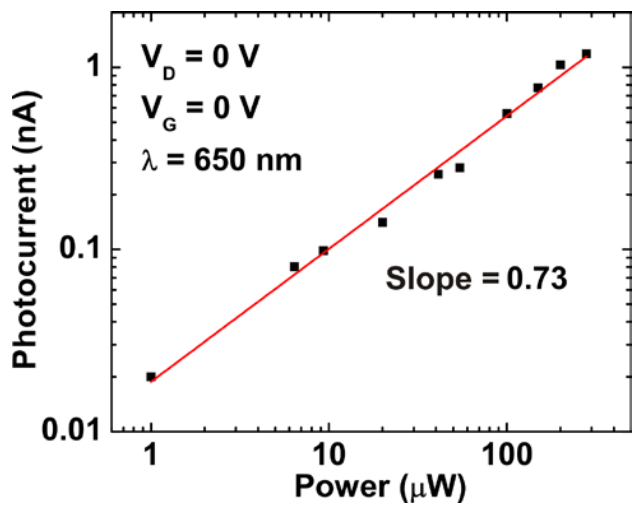

**Figure S8:** Power dependence of the photocurrent under zero applied bias. The red line is a linear fit to the data showing a slope of 0.73.

**b. Calculation of Responsivity and Quantum Efficiency**

The quantum efficiency was calculated using the formula $EQE = \dfrac{I_{ph}/e}{P/h\nu} \times 100$, where $I_{ph}$, e, P, and hν represent the photocurrent, electronic charge, incident optical power, and photon energy, respectively. The photocurrent was extracted by subtracting the dark I-V curves from the illuminated I-V curves. The responsivity (R) was calculated using $R = EQE \times 1.24/\lambda$, where λ is in μm.